\documentclass[12pt,titlepage]{article}
\usepackage{epsfig}
\usepackage{epsf}
\usepackage{color}
\usepackage{graphicx}
\usepackage{amssymb}
\def\beq{\begin{equation}}
\def\eeq{\end{equation}}
\def\bea{\begin{eqnarray}}
\def\eea{\end{eqnarray}}
\begin{document}
\title{Pressure and volume in the first law of black hole thermodynamics}

\author{{Brian P. Dolan}\\
{\small Department of Mathematical Physics, National University
of Ireland,}\\
{\small  Maynooth, Ireland}\\
{\small and}\\
{\small Dublin Institute for Advanced Studies,
10 Burlington Rd., Dublin, Ireland}\\
{\small e-mail: bdolan@thphys.nuim.ie}}

\maketitle
\begin{abstract}
The mass of a black hole is interpreted, in terms of thermodynamic 
potentials, as being the
enthalpy, with the pressure given by the cosmological constant.
The volume is then defined as being the Legendre transform of the 
pressure and the resulting relation between volume and pressure is
explored in the case of positive pressure.  A virial expansion is developed and a van der Waals like
critical point determined. 
The first law of black hole thermodynamics includes
a $PdV$ term which modifies the maximal efficiency of a Penrose
process.  It is shown that, in four dimensional space-time
with a negative cosmological constant,
an extremal charged rotating black hole can have an efficiency of up
to 75\%, while for an electrically neutral rotating back hole this
figure is reduced to 52\%, compared to the corresponding values of 50\% and 
29\% respectively when the cosmological constant is zero.

\bigskip

\ 

\noindent
\leftline{\hbox{PACS nos: 04.60.-m;  04.70.Dy}}
\end{abstract}

\section{Introduction}

The thermodynamics of black holes is a rich and fascinating area of research
which continues to yield surprises.
The first law of black hole thermodynamics is usually written as
\beq \label{firstlaw}
d M= TdS + \Omega d J + \Phi d Q 
\eeq
where $T=\frac{\kappa}{2\pi}$ is the Hawking temperature of the black hole
(with $\kappa$ the surface gravity), $S=\frac A 4$ the entropy
(with $A$ the area in Planck units),
$\Omega$ the angular velocity, $J$ the angular momentum, $\Phi$ the electrostatic potential difference between infinity and the horizon, 
$Q$ the electric charge and $M$ the mass. The mass is usually
interpreted as the internal energy, in the thermodynamic sense, 
of the black hole, but it was suggested
in \cite{KRT} that it is more correctly interpreted as the enthalpy.
In this context it is notable that there is no $PdV$ term in (\ref{firstlaw}), corresponding to a change in volume at ambient pressure $P$. When a cosmological
constant, $\Lambda$, is included there is a natural candidate for a
pressure, $P=-\frac{\Lambda}{8\pi}$, and it was proposed in \cite{Enthalpy}
that the volume of the black
hole be defined as the thermodynamic variable conjugate to $P$.
Interpreting the mass as the enthalpy, equation (\ref{firstlaw}) should then be  modified to
\beq \label{dM}
d M= TdS  +V d P+\Omega d J + \Phi d Q,
\eeq
where the thermodynamic volume is defined to be $V=\left.\frac{\partial M}{\partial P}\right|_{S,J,Q}$, \cite{KRT}. 
The idea that $\Lambda$ should be thought of as a thermodynamic variable that can be varied is not new and has been considered by a number of authors, \cite{HT}-\cite{CCK}.
Equation (\ref{dM}), with $Q=0$, was studied in \cite{Wangetal}, in the context
of varying $\Lambda$, without any particular physical interpretation being given
to the thermodynamic conjugate of $\Lambda$, often denoted $\Theta$.  

One may question whether it is appropriate to identify $\Lambda$ with a thermodynamic pressure.  While a cosmological constant gives a pressure term in Einstein's equations, a fluid dynamical pressure
is not necessarily the same as a thermodynamic pressure.
In equilibrium situations however it is presumably correct to identify 
the fluid dynamical pressure with the thermodynamic
pressure and we shall do so here.   

In general the enthalpy, $H$, 
is the heat energy
beloved of chemists, it is not the internal energy, $U$, of the first law
of thermodynamics.  That distinction goes to the Legendre transform of the
enthalpy,
\beq U=H-PV,\eeq
where $H$ is a function of $S$, $P$, $J$ and $Q$
while $U(S,V,J,Q)$ is a function of purely extensive variables.
Then we get the usual form of the first law,
\beq \label{FirstLawIntro}
d U= TdS  - PdV +\Omega d J + \Phi d Q.
\eeq
This equation was written down in \cite{CCK}, using $\Theta-\Lambda$ notation, but its consequences were not pursued.

When $\Lambda$ is non-zero we should expect the $PdV$ term
to contribute to the mechanical energy that can be extracted from a
black hole, by a Penrose process for example. 
For a negative $\Lambda$
(positive pressure) the $PdV$ term gives a positive contribution to $dU$ if the black hole shrinks, and 
the $PdV$ term reduces the amount of energy available for extraction as mechanical work $W$, with $dW=-dU$, hence reducing the efficiency. However we shall show that the maximal efficiency 
actually increases, 
relative to the $\Lambda=0$ case, when $\Lambda<0$, because the
maximal angular momentum of a black hole in AdS is greater
than that of one with $\Lambda=0$ and this can outweigh
the reduction in efficiency: the engine may not be as efficient
at a given $J$ but it can be pushed to higher $J$.
Conversely one would expect that, for a positive $\Lambda$ more energy becomes
available at a given $J$, relative to $\Lambda=0$, as the black hole shrinks.

Of course there are no pistons pushing against a gas 
for a black hole, but a negative cosmological
constant contributes a negative energy density
to space-time so a shrinking black-hole exposes negative
energy, thus increasing the black hole's  internal energy
and decreasing the amount of energy available
for mechanical work. Conversely a positive cosmological 
constant would presumably release extra energy as the black hole shrinks, that can be used to do work, hence increasing the efficiency at a given $J$.

The most efficient way to extract energy from a black hole
is in an isentropic process, with $dS=0$ and the area
of the event horizon constant.  So
\beq dU \ge dU_{min}= -P dV +\Omega dJ + \Phi d Q.\eeq
We shall see that, for a rotating black hole, it is possible to
reduce $V$ while keeping $S$ constant.
The maximum amount of mechanical work that
can be extracted in passing from an initial state $i$
to a final state $f$ is
\beq W_{max}=-\int_i^f dU_{min}. \eeq
The efficiency is defined to be
the ratio of the mechanical energy extracted to the initial heat energy (enthalpy),
\beq \eta=\frac {W_{max}}{M_i},
\eeq
where the initial enthalpy is identified with the initial mass, $M_i$.
It will be shown that, for $\Lambda<0$, this can be as high as 52\% for a 
rotating neutral black hole and 75\% for a charged black hole.

The volume of a black hole has only recently
been considered as a thermodynamic variable, \cite{Enthalpy,CGKP}. 
At the simplest level, there is a natural tendency to 
assume that the area and the volume are related geometrically 
and are not independent.  
For a Schwarzschild black hole with radius $r_h$, for example, the area is of course $4\pi r_h^2$ and indeed the thermodynamic volume works out to be $\frac{4\pi }{3}r_h^3$, but this seems co-incidental and no particular significance should
be attached to it.
At a deeper level it is not even clear how
to define a volume geometrically, as the metric is not static in
the interior of a black hole.
For a Schwarzschild black hole which is not rotating we shall see that
the thermodynamic volume and the area are not independent: fixing $S$ fixes $V$ so $dV=0$ in an isentropic process, and the $PdV$ term does not contribute to the first law.  But for a rotating black hole the area
of the event horizon does not determine the thermodynamic volume uniquely and it is possible to vary the volume
keeping the entropy constant, by changing the angular momentum and/or the charge.

 The properties of the thermodynamic volume and its contribution to the first law of black hole thermodynamics are
explored in detail in this paper for a rotating
charged black hole in four dimensional space-time with a negative
$\Lambda$.
In section \ref{S-InternalEnergy}  thermodynamic potentials and the equation of state are discussed and the Legendre
transform from the enthalpy to the internal energy is given explicitly. 
In section \ref{JQ-holes} the efficiency of a Penrose type process is analysed and section \ref{Conclusions} contains a discussion and outlook. Two appendices are dedicated to the technicalities of deriving some results used in the text

\section{The  internal energy}\label{S-InternalEnergy}

Including a pressure term in the first law gives
\[ dU = TdS -PdV + \Omega d J + \Phi d Q\]
where the internal energy, $U(S,V,J,Q)$, is a function of extensive variables.
The thermodynamic volume $V$ is the conjugate variable to the pressure and is
obtained from the mass, which is identified in \cite{KRT} with the enthalpy,
$M=H(S,P,J,Q)$,
by 
\beq \label{ThermodynamicVolume}
V=\left.\frac{\partial H}{\partial P}\right|_{S,J,Q}.\eeq
 It was proposed in \cite{Enthalpy} that (\ref{ThermodynamicVolume})
be defined to be the thermodynamic volume of the black hole.

The line element for a charged rotating black hole in 4-dimensional anti-de Sitter 
space is \cite{Carter}
\beq  \label{ChargedAdSKerr}
d s^2=-\frac{\Delta}{\rho^2}\left( d t - \frac{a\sin^2\theta}{\Xi}\, d\phi \right)^2
+\frac{\rho^2}{\Delta}dr^2 + \frac{\rho^2}{\Delta_\theta}d\theta^2 
+\frac{\Delta_\theta\sin^2\theta}{\rho^2}\left(a d t - \frac{r^2 + a^2}{\Xi}\,d\phi \right)^2,
\eeq
where 
\bea \label{MetricFunctions}
\Delta&=&\frac{(r^2+a^2)(L^2+r^2)}{L^2} -2 m r + q^2, \qquad
\Delta_\theta=1-\frac{a^2}{L^2}\cos^2\theta,\nonumber \\
\rho^2& = & r^2 + a^2\cos^2\theta, \qquad \Xi = 1-\frac{a^2}{L^2},
\eea
and the cosmological constant is $\Lambda=-\frac 3 {L^2}$,
which is related to the pressure by $\frac {1} {L^2} = \frac {8\pi P}{3}$.

The physical properties of this space-time are well known \cite{KNAdS}, and the first law, applied to this metric, was discussed in \cite{GPP}, but without a $PdV$ term.
The metric parameters $m$ and $q$ are related to the mass and charge by
\beq \label{MassCharge}
M=\frac{m}{\Xi^2},\qquad Q=\frac {q} {\Xi}.\eeq
The event horizon, $r_+$, lies at the largest root of $\Delta(r)=0$, so
\beq \label{Mass}
M=\frac{(r_+^2+a^2)(L^2+r_+^2) + q^2 L^2}{2 r_+ L^2 \,\Xi^2}, \eeq
and the area of the event horizon is
\beq \label{Area}
A=4\pi\frac{(r_+^2+a^2)}{\Xi}.
\eeq
The temperature is 
\beq \label{Temperature}
T=\frac{(L^2+3 r_+^2)r_+^2-a^2(L^2-r_+^2)-q^2 L^2}{4\pi L^2 r_+
(r_+^2+a^2) }.\eeq

The angular momentum, $J=a M$ and the relevant thermodynamic
angular velocity is
\beq
\Omega=\frac{a(L^2+r_+^2)}{L^2(r_+^2+a^2)}.   
 \eeq
The electrostatic potential is
\beq
\Phi=\frac{q r_+}{r_+^2+a^2}.
\eeq

One can scale $L$ out from all the above expressions
by defining dimensionless variables $\overline M = M/L$, $\bar a = a/L$, $\bar r_+ = r_+/L$, etc., but we prefer to keep $L$ explicit to expose more clearly the r\^ole of the pressure, and make the 
comparison with the $L\rightarrow\infty$ limit clear.

Under the assumptions made here the thermodynamic volume for the Kerr-Newman-AdS black hole works out to be
\beq \label{Volume}
V = \frac {2\pi }{3}\left\{
\frac{(r_+^2+a^2)\bigl(2 r_+^2 L^2 + a^2 L^2 - r_+^2 a^2 \bigr) +  L^2 q^2 a^2}
{L^2 \Xi^2 \,r_+}\right\},
\eeq
which is a simple generalisation of the Kerr-AdS volume
derived in \cite{CGKP}.  A direct derivation of (\ref{Volume})
from (\ref{Mass}) is most easily achieved by first writing the
mass as a function of $(S,P,J,Q)$, differentiating with respect 
to $P$, and then transforming back to $(r_+,a,q,L)$.

When $a=0$ the area (\ref{Area}) and volume (\ref{Volume}) are not independent and
the area determines the volume uniquely, but when $a$ is
non-zero the area and the volume become independent.
 
For asymptotically flat space, with $L\rightarrow\infty$, 
one has
\beq 
V = \frac {2\pi }{3}
\frac{(r_+^2+a^2)\bigl(2 r_+^2  + a^2   \bigr) +q^2a^2}
{r_+}.
\eeq
While this reduces to the na{\"\i}ve result, $\frac{4\pi r_+^3}{3}$, 
for the Schwarzschild black hole, a geometrical interpretation when $a$ 
is non-zero is not so clear.

A correct description of the thermodynamics of the black hole,
in terms of thermodynamic potentials, requires replacing
the geometric variables $(r_+,L,a,q)$ with thermodynamic
variables $(S,P,J,Q)$.  The relevant expression for the
mass, and hence the enthalpy, was derived in \cite{CCK},
\beq \label{CCKmass}
H(S,P,J,Q):=
\frac {1}{2}\sqrt{\frac{
\left(S   + \pi Q^2 + \frac{8 P S^2}{3}\right)^2 + 
4 \pi^2\left( 1+\frac{8 P S}{3}  \right) J^2}
{\pi S}}.
\eeq
This generalises the Christodoulou-Ruffini formula \cite{CR} for the mass
of a rotating black hole in terms of its irreducible mass, $M_{irr}$.  The irreducible mass for a black hole with entropy $S$
is the mass of a Schwarzschild black hole with the same 
entropy, $M^2_{irr}=\frac {S}{4\pi}$.

In terms of thermodynamic variables the temperature is, \cite{CCK},
\beq \label{CCKTemperature}
T= \left.\frac {\partial H}{\partial S}\right|_{J,Q,P}=\frac {1}{8\pi H}\left[
\left(1  +\frac{\pi Q^2}{S}+\frac {8 P S}{3} \right)
\left(1 - \frac{\pi Q^2}{S}+ 8 P S \right)
-4\pi^2 \left(\frac {J}{S}\right)^2\right]
\eeq
and the thermodynamic volume is
\beq
V= \left.\frac {\partial H}{\partial P}\right|_{S,J,Q}=\frac{2}{3 \pi H}\left[S\left(S + \pi Q^2 + \frac{8PS^2}{3}\right)+ 2\pi^2 J^2  \right],
\eeq
which is manifestly positive.

The Legendre transform $U=H-PV$ gives the thermal energy,
a function of purely extensive variables.  The transform is 
evaluated in an appendix to be
\bea \label{InternalEnergy}
U(S,V,J,Q)&=&  \left(\frac {\pi}{S}\right)^3\left[ 
\left(\frac{3V}{4\pi}\right)
\left\{ \left(\frac S {2\pi}\right)\left(\frac S {\pi} + Q^2 
\right)+J^2 \right\}\vline  height 20pt  width 0pt depth 20pt
\right.\\
& & \kern 50pt \left. -|J|\left\{\left(\frac{3V}{4\pi}\right)^2-\left(\frac S \pi \right)^3\right\}^{\frac 1 2}
\left(\frac{S Q^2}{\pi}+J^2\right)^{\frac 1 2}
 \right].  \nonumber
\eea
Dimensional analysis implies that $U$ is only a function
of three independent variables, since $U\rightarrow \lambda U$ when $S\rightarrow \lambda^2 S$, $V\rightarrow \lambda^3 V$, $J\rightarrow \lambda^2 J$ and
$Q\rightarrow \lambda Q$.

One must be careful taking the $J\rightarrow 0$ limit of these potentials. In this limit the enthalpy
\beq \label{ZeroJEnthalpy}
H(S,P,0,Q)=\frac 1 2 \sqrt{\frac S \pi} \left( 1+ \frac {\pi Q^2}{S} +\frac{8SP}{3} \right)
\eeq
is linear in $P$ and the Legendre transform is singular: 
\beq
V=\left.\frac {\partial H}{\partial P}\right|_{S,Q}=
\frac{4\pi}{3}\left(\frac {S}{\pi} \right)^{3/2}\eeq
is independent of $P$ and so cannot be inverted to obtain  $P(V)$. Conversely, when $J$ is zero, the Legendre transform of  (\ref{ZeroJEnthalpy}) is
\beq 
U=\frac 1 2 \sqrt{\frac S \pi} \left( 1+ \frac {\pi Q^2}{S} \right), \label{ZeroJEnergy}
\eeq
which is independent of the volume and is not equal to the
$J\rightarrow 0$ limit of (\ref{InternalEnergy}),
unless a constraint, $\bigl(\frac{3V}{4\pi}\bigr)^2=\left(\frac S \pi \right)^3 $, is imposed.  Indeed (\ref{ZeroJEnergy}) gives
the  wrong $J=0$ temperature (unless $Q=P=0$, in which case 
$U=H$ and $T=\frac{\partial U}{\partial S}=\frac{1}{4\pi r_+}$
is the correct Hawking temperature for a 
Schwarzschild black hole). 
To get the correct temperature from (\ref{InternalEnergy})
in the $J\rightarrow 0 $ limit 
we must take the partial derivative with respect to $S$
{\it before} setting $J$ to zero, and take note of
the fact that $\bigl(\frac{3V}{4\pi}\bigr)^2=\left(\frac S \pi \right)^3 $ when $J=0$. 
This constraint can be derived from (\ref{InternalEnergy}) directly by observing that $U$ is not differentiable with
respect to $J$ at $J=0$ unless
$\Bigl(\bigl(\frac{3V}{4\pi}\bigr)^2-\bigl(\frac S \pi \bigr)^3\Bigr)\Bigl(J^2 + \frac {S Q^2}{\pi}\Bigr) $ vanishes there.
It can also be seen directly when $a=0$ in 
(\ref{Area}) and (\ref{Volume}).

To derive the relation between the pressure and the volume in general
we first define $v:=\frac{3 V}{4\pi}$ and $s:=\frac S \pi$. Then (\ref{InternalEnergy}) becomes
\beq \label{internalenergy}
U(s,v,J,Q)= 
\frac{1}{s^3}\left\{ \frac v 2
\bigl(s^2 + s\, Q^2  + 2 J^2 \bigr)
-|J|\sqrt{\left(v^2-s^3\right)
\left(J^2 + s\, Q^2\right) 
} \right\}
\eeq
with temperature
\beq \label{Utemp}
T=\frac {1} {\pi} \frac{\partial U}{\partial s}
=\frac{|J|
\left\{3 J^2\bigl(2v^2 - s^3\bigr) +s\, Q^2\bigl(5v^2-2s^3\bigr)\right\}}
{2\pi s^4 \sqrt{\bigl(v^2-s^3 \bigr)\bigl(J^2 + s\, Q^2 \bigr) }}
-\frac{v\Bigl(s^2 + 2 s\, Q^2 + 6 J^2 \Bigr)}{2 \pi s^4}
\eeq
and pressure
\beq \label{UP}
P=-\frac {3} {4\pi} \frac{\partial U}{\partial v}
= \frac{3 v|J|}{4\pi s^3} 
\sqrt{\frac{J^2 + s\, Q^2}{v^2-s^3}}-\frac{3}{8\pi s^3} \left(s^2 + s\, Q^2 +2J^2 \right).
\eeq
In the $J\rightarrow 0$ limit
$|J|$ and $\sqrt{v^2-s^3}$ must vanish together
for finite $T$ and $P$.
The equation of state, in the form of the relation between the pressure, the temperature and the volume, can be obtained by eliminating 
$s$ between (\ref{Utemp}) and (\ref{UP}).  
In the limit $J\rightarrow 0$ the temperature is
\beq \label{EnthalpyTemp}
T_{J\rightarrow \,0}=
\frac {8\pi v^{ 4/3}P+ v^{2/3}-Q^2}{4 \pi v },
\eeq
which is the correct temperature as a function of
$P$, $Q$ and $s$, as can be seen from (\ref{Temperature}), with $a=0$, using 
(\ref{MassCharge}), (\ref{Area}), (\ref{Volume}) and $J=aM$.  When $Q=0$
(\ref{EnthalpyTemp}) reduces to the equation of state in \cite{Enthalpy}.

The small $J$ correction to (\ref{EnthalpyTemp}) is
\beq
T=\frac{8\pi P v^2 + v^{4/3} - Q^2 v^{2/3} - 6 J^2 }{4\pi v^{5/3}} + \ 
O\left(\frac {J^4}{v^{9/3}},\frac{J^2 Q^2}{v^{9/3}}\right).
\eeq
The angular momentum term dominates at small volumes for any $J>0$.
The first few  terms in an expansion for the pressure are
\beq  \label{PVapprox}
P=\frac T {2v^{1/3}} - \frac 1 {8\pi v^{2/3}} + \frac {Q^2}{8\pi v^{4/3}} + \frac{3 J^2}{4 \pi v^2}
+\ O\left(\frac{J^4}{v^{10/3}},\frac{J^2 Q^2}{v^{10/3}} \right),
\eeq
(the virial expansion for $Q=0$ is developed in more detail in the second appendix.)

The $P$-$V$ diagram is plotted to this order
in figure 1 with $Q=0$ and $J=1$ (for clarity the horizontal axis in figure 1 is
$v^{1/3}$ rather than $v$).  There is a critical point at
$T_c\approx 0.0413$, $P_c\approx 0.00280$ and $v_c^{1/3}\approx 3.08$.
This critical point represents a second order phase transition, similar
in nature to that of a van der Waals gas --- it
is the same critical point that was found in \cite{CCK} by fixing the
pressure and varying $J$ (the authors of \cite{CCK} use 
$P=\frac {3}{8\pi}=0.119$, corresponding to $L=1$, and find a critical 
value of $J$ at $J_c=0.0236$ giving $\frac {3 J_c}{8 \pi}=0.00282$).
The corrections in equation (\ref{PVapprox}) become large
when $v$ is very small,
but at the critical point the error in the pressure is of order $\sim \frac 1 {v^{10/3}}\sim 10^{-5}$, or about 1\% of $P_c$.  
When $v^{1/3} \lesssim 2$ the corrections are of the
same order as $P$ and the virial expansion breaks down.
In the limit when $v$ is small relative to $J$, a large $J$ expansion shows
that the  leading order behaviour is $P\approx \frac{3 J^2}{8\pi v^2}$, which is the same low $v$ behaviour as figure 1
but reduced by a factor 2.

Thus by taking the $J\rightarrow 0$ limit carefully the
temperature can be kept  finite when $v^2\rightarrow s^2$, but the non-analyticity in (\ref{InternalEnergy}) is still present
in the second derivative with respect to $S$, which diverges when $J=0$ and $v^2=s^3$.  This has the important
consequence that the heat capacity at constant volume, 
$C_V=T/
\left(\frac {\partial T} {\partial S} \right)_{V,J,Q}$, 
tends to zero when $J=0$, as observed in \cite{Enthalpy}.
$C_V$ can be  non-zero for $J\ne 0$, and it is plotted in figure 2 for $Q=0$ as a function of $S$ and $J$, with $V=1$. 
For comparison the heat capacity at constant pressure, 
$C_P=T/
\left(\frac {\partial T} {\partial S} \right)_{P,J,Q}$,
is plotted in figure 3 for $Q=0$ as a function of
$S$ and $J$, with $P=1$. 
$C_P$ vanishes when $T=0$ and diverges when $\frac {\partial T} {\partial S}$ =0. 

A full stability analysis was carried out in \cite{CCK} and figure 4 shows
the phase diagram, when $Q=0$, in terms of $S/L^2$ and $J/L^2$.
The top curve, labelled I,  is the $T=0$ limit, the region above and to the left of this curve is unphysical because $T<0$ there.
$C_P$ in figure 3 vanishes on curve I.  Curve II is the curve on which $\Omega=1/L$, below which the black hole can be in equilibrium with radiation rotating at infinity \cite{HHTR}. On this curve asymptotic space-time is conformal to the $2+1$ dimensional Einstein universe rotating at the speed of light, the region below and to the right of this curve is of relevance to conformal field theories. Curve III is the curve marking the edge of local stability, the determinant of the Hessian of the Gibbs free energy is divergent on this curve.  
The heat capacity, $C_P$, diverges on curve IV,
which is blown up in the right hand picture (this curve was not included in the
analysis of reference \cite{CCK}).
Curve V marks the Hawking-Page phase transition, below this curve the black hole is globally stable. 
The three curves marking the Hawking-Page transition, the edge
of local stability and the $\Omega=1/L$ curve all have similar
asymptotic forms for large $S/L^2$, $\frac J {L^2} \approx \frac {S^2}{2 \pi^2 L^4}
+ O\left(\frac{S}{L^2}\right)$.

In the next section 
it will be more convenient to use the variables 
$S$, $P$, $J$ and $Q$, rather than $S$, $V$, $J$ and $Q$,
in terms of which 
\beq \label{USPJQ}
U= \frac{\Bigl( S   + \pi Q^2\Bigr)
\left(S   + \pi Q^2 + \frac{8 P S^2 }{3} \right)
+ 4\pi^2 \left(1 + \frac{4 P S} {3} \right) J^2 }
{2\sqrt{ \pi S \left[
\left(S   + \pi Q^2 + \frac{8 P S^2}{3}\right)^2 + 
4 \pi^2\left( 1+\frac{8 P S}{3}  \right) 
J^2\right]}},
\eeq
which is manifestly positive.
In terms of geometrical variables,
\beq \label{ChargedInternalEnergy}
U= \frac{(r_+^2+a^2)\bigl(2L^4+a^2r_+^2-a^2L^2)
+q^2 L^2 (2L^2 - a^2\bigr)}{4L^4\,\Xi^2\, r_+}.
\eeq
\section{Extracting energy from rotating black holes}\label{JQ-holes}

For an electrically
neutral black hole, we set $Q=0$ in (\ref{USPJQ}) and
the internal energy $U=H - PV$, written in terms of $S$, 
$L^2=\frac {3}{8\pi P}$ and $J$, is  
\beq \label{USPJ}
U= \frac{
S^2\left(1   +  \frac{S}{\pi L^2} \right)
+ 4\pi^2  J^2 \left(1 + \frac{S} {2 \pi L^2} \right)}
{2\sqrt{\pi S \left(1 + \frac{S} {\pi L^2} \right) \left[
S^2\left(1   + \frac{S}{\pi L^2}\right) + 
4 \pi^2
J^2\right]}} 
\;.
\eeq
In an isentropic, isobaric process the black 
hole can 
yield mechanical work 
by decreasing the angular momentum.  If $J$ is reduced from some finite value to zero the efficiency is
\bea \eta&=& \frac{U(J) - U(0)}{H(J)} \nonumber \\
&=& \frac {S^2\left(1 + \frac{S}{\pi L^2}\right) 
+4\pi^2\left(1+\frac{S}{2\pi L^2}\right)J^2}
{\left(1+\frac{S}{\pi L^2}\right)
\left[S^2\left(1+\frac{S}{\pi L^2}\right)
+4\pi^2 J^2\right]}\\
&&\hskip 2cm-\frac{S}{\sqrt{
\left(1+\frac{S}{\pi L^2}\right)
\left[S^2\left(1+\frac{S}{\pi L^2}\right)
+4\pi^2 J^2\right]
}}\;.
\nonumber
\eea

For a given $S$ and $L$ there is
a maximal value of $J$ determined by demanding that $T>0$,
which requires 
\beq \label{JExtremalQzero}
J^2<\frac{S^2}{4\pi^2}\left(1+ \frac{S}{\pi L^2} \right)\left(1+\frac{3S}{\pi L^2} \right).
\eeq
The greatest efficiency is for extremal black holes,
when the bound (\ref{JExtremalQzero})
is saturated,
\beq \eta=\frac{\left(4\pi L^2+3 S\right)} 
{2\left(2\pi L^2 +3 S\right)}-
\frac {\pi L^2}{\left(\pi L^2+S\right)\sqrt{2\pi L^2+3S}}.
\eeq
The asymptotically flat case ($\Lambda=0$) is obtained by taking the limit $L\rightarrow\infty$,
$\eta=1-\frac{1}{\sqrt{2}}=0.29\ldots$, 
which is the familiar result \cite{Wald}.
For finite $L$ the efficiency is greater than the asymptotically flat value.
It has a maximum value of 
$0.5184...$
when $S/L^2=15.39...$ (obtained by solving a quartic equation for $S$), 
and asymptotes to $0.5$ for large $S$.

Differentiating
(\ref{CCKmass}) with respect to $P$, we see that the thermodynamic volume  is a monotonically
increasing function of $J$, hence the volume decreases
as $J$ is lowered, keeping the area constant.
The black hole also heats up during the process.

When the black hole is charged there are various possibilities for extracting work
through an  isentropic, isobaric process: one could decrease $J$ keeping $Q$ constant
or decrease $Q$ keeping $J$ constant or decrease both simultaneously.  Since all that matters is the initial and final values of $U$, the initial and final values of $J$ and $Q$ uniquely determine the efficiency in such a process.

If both $J$ and $Q$ are both  non-zero initially and
are both decreased to zero, the efficiency is 
\bea \eta \label{Qefficiency}
&=& \frac {(S+\pi Q^2)\left(S + \pi Q^2+\frac{S^2}{\pi L^2}\right) 
+4\pi^2\left(1+\frac{S}{2\pi L^2}\right)J^2}
{\left(S+ \pi Q^2+\frac{S^2}{\pi L^2}\right)^2
+4\pi^2\left(1+\frac{S}{\pi L^2}\right)J^2}\\
&& \kern 60pt -\frac{S}{\sqrt{\left(S+\pi Q^2+\frac{S^2}{\pi L^2}\right)^2+
4\pi^2\left(1+\frac {S}{\pi L^2}\right)J^2}}\;.
\nonumber
\eea
The optimal efficiency is for  extremal black holes 
and the extremal value of $J$ is given by
\beq \label{QJmax}
J_{max}^2=\frac {1} {4}\left[
\left(\frac {S} {\pi}\right)^2
\left( 1+\frac{S}{\pi L^2}  \right) 
\left( 1+\frac{3S}{\pi L^2}  \right) 
+ 2\left(\frac {S} {\pi}\right)^2\left(\frac{Q}{L}\right)^2- Q^4\right],
\eeq
when $T$ in (\ref{CCKTemperature}) vanishes.
Positivity of $J_{max}^2$ then imposes a restriction on the allowed range of $Q^2$, 
\beq \label{Qrange}
0\le Q^2 \le 
\left(\frac {S} {\pi}\right) \left( 1+\frac{3S}{\pi L^2}  \right).
\eeq
The greatest efficiency is for an extremal black hole with maximal charge, 
\beq Q_{max}^2 =
\left(\frac {S} {\pi}\right) \left( 1+\frac{3S}{\pi L^2}  \right),
\eeq
when
\beq \eta=\frac{\pi L^2 +3S}{2\bigl(\pi L^2 +2S\bigr)}\;.
\eeq

For large $S$ efficiencies of up to $75\%$ are possible
in principle for extremal charged, rotating AdS black holes, though as
we saw in the previous section
this is reduced to $51.8\%$ for extremal rotating, but electrically neutral, black holes.  For comparison, the equivalent figures in asymptotically flat space-time, with $\Lambda=0$, are $50\%$ and $29.3\%$ respectively. 

\section{Conclusions}\label{Conclusions}

The thermodynamics of rotating black holes in
four dimensional in anti-de Sitter space-time 
has been discussed in detail, with particular attention
payed to the r\^ole of pressure and the
volume.  The negative cosmological 
constant is interpreted as a positive pressure and treated
as a thermodynamic variable whose conjugate thermodynamic
variable is a thermodynamic volume.  
For rotating black holes the thermodynamic volume
is independent of the geometric area, and hence independent
of the entropy, allowing a full thermodynamic treatment.

The black hole mass is associated with the enthalpy $H$,
as suggested in \cite{KRT},
and the internal energy $U=H-PV$ is lowered relative
to the mass by a positive pressure.  Vanishing angular
momentum is a non-analytic point of the thermodynamics and
must be treated with some care, in particular one cannot fully
understand black hole thermodynamics simply by focusing on
the non-rotating, Schwarzschild case --- rotation
is essential for a complete analysis.

The black hole equation of state has been analysed 
in terms of pressure and volume and a virial expansion
developed.  Non-zero angular momentum causes
a rapid rise in pressure as the volume is reduced
at constant temperature, a rise that is not present
at zero angular momentum.  The critical point
found in \cite{CCK} is found to be very similar in form 
to that of a van der Waals gas.

In a Penrose process the volume decreases as the black
hole looses angular momentum and the $PdV$ term in the
first law reduces the amount of energy available to
do work.  However a negative cosmological constant also
extends the range of the angular momentum, pushing
the maximum allowed value beyond
that of the $\Lambda=0$ case, so that in fact more work can be 
extracted from an extremal black hole in a space-time that
is asymptotically anti-de Sitter than from one in a space-time that is asymptotically flat.  Efficiencies of up to 75\%
are theoretically achievable for a charged black hole in
asymptotically anti-de Sitter space-time, compared to 50\%
in asymptotically flat space-time.  For electrically
neutral black holes the corresponding figures are 51.8\%
compared to 29.3\%.   

The analysis here has focused on the $\Lambda<0$ case
for three reasons.  Firstly there is an ambiguity in 
the definition of thermodynamic variables in the $\Lambda>0$
case, there are two relevant event horizons to be considered
and hence two different temperatures in general.
For a general choice of black hole parameters it is not
possible to make the Euclidean geometry regular by
a unique periodic identification of Euclidean time.
Secondly $\Lambda>0$ implies negative pressure and 
consequent thermodynamic 
instabilities.  
This latter problem
is not necessarily a fatal objection, black holes are well
known to be unstable for $\Lambda=0$ because the heat
capacity, more precisely $C_P$, is negative, but we
can still consider metastable situations in which 
a temperature can be defined for a period of time much
shorter than the timescale of thermal instability.
Indeed there are regions of parameter space for $\Lambda<0$
in which $C_P<0$, and we can still learn something about
physical properties of black holes in this regime.  Negative
pressures can be useful in metastable situations \cite{LL}.
Lastly it is not even clear how to define the mass of a black hole
in a $\Lambda>0$ space-time as there is no accessible asymptotic regime
in which one can compare with the zero mass solution.

\appendix

\section*{Appendix 1}

To calculate the internal energy $U=H-PV$, we write
the enthalpy in the form
\beq H=\sqrt{\alpha + \beta P +\gamma P^2},
\eeq
where
\bea 
\alpha&:=& \frac{\pi}{S}\left\{ \frac 1 4 \left( \frac S \pi + Q^2 \right)^2 + J^2 \right\} \nonumber \\
\beta &:=& \frac {4\pi}{3}\left\{ \frac S \pi \left( \frac S \pi + Q^2 \right)+2J^2 \right\}
\\
\gamma &:=& \left(\frac{4\pi}{3}\right)^2\left(\frac{S}{\pi}\right)^3.\nonumber 
\eea
Note that the discriminant,
\beq \beta^2 - 4\alpha\gamma = \frac{64 \pi^2}{9} J^2 
\left(J^2 +\frac {S Q^2}{\pi}\right), 
\eeq
is positive.

Now
\beq V=\left.\frac {\partial H}{\partial P}\right|_{S,J,Q}
= \frac{\beta + 2 \gamma P}{2 H}
\qquad \Rightarrow \qquad P=\frac{2 H V - \beta}{2 \gamma}.
\eeq
This allows us to re-express $H$ as a function of $V$,
\beq H=\frac 1 2 \sqrt{\frac {\beta^2 - 4\alpha\gamma}
{V^2 - \gamma}}.
\eeq
We can immediately conclude that 
\beq
V^2>\left(\frac{4\pi}{3}\right)^2\left(\frac{S}{\pi}\right)^3,
\eeq
in agreement with the observation in \cite{CGKP}.

It is now straightforward to determine
\beq
U=H-PV= H- \left(\frac {H V^2}{\gamma} - \frac{\beta V}{2\gamma}\right)
=  \frac{\beta V}{2\gamma} - 
\frac {\sqrt{\bigl(V^2 - \gamma\bigr) \bigl(\beta^2-4\alpha\gamma\bigr)}}{2\gamma},
\eeq
which immediately gives (\ref{InternalEnergy}) in the text.

\section*{Appendix 2}

In this appendix the virial expansion is developed.  For simplicity
we set $Q=0$, but the same techniques can be applied 
to the case of non-zero $Q$.
To develop the expansion we use dimensionless variables
\beq
y:=\frac {v} {J^{3/2}},\quad x:=\frac {s} {J}, 
\quad
p:= \frac {8\pi P J}{3}, \quad t:= 2\pi T J^{1/2},
\eeq
in terms of which equations (\ref{Utemp}) and (\ref{UP}) can be written
\bea  \label{tequation}
t&=&\frac{3(2 y^2 - x^3)}{x^4\sqrt{y^2-x^3}} - \frac {y(x^2+6)}{x^4}
\\
p&=&\frac {2 y}{x^3\sqrt{y^2-x^3}} -\frac{x^2+2}{x^3}.
 \label{pequation}
\eea
When $y$ and $x$ are large let $y^2\approx x^3$, with $y^2-x^3=z^2$, 
then $t$ is
finite provided  $z\approx \frac{3}{y^{2/3} t}$ in which case
$p\approx \frac {2 t} {3 y^{1/3}}$.  
Replacing $x$ with $z$ in equations (\ref{tequation}) and (\ref{pequation}) gives
\bea \label{t-z}
3\, \left( {y}^{2}+z^2 \right)&=&\left\{ t \left( {y}^{2}-
z^2 \right) ^{4/3}+ y \left( {y}^{2}-z^2 \right) ^{2/3}+6\,y
 \right\}z,\\
\label{p-z}
(y^2-z^2)z p &=& 2y - z\Bigl\{ (y^2-z^2)^{2/3}+2\Bigr\}\nonumber\\
\Rightarrow \qquad
p&=&\frac {2}{z(y+z)}-\frac{1}{(y^2-z^2)^{1/3}}.
\eea
Now we expand in powers of $u=\frac {1}{y^{1/3}}$.
Let $z=\frac{3 u^2}{t}\zeta$ for some $\zeta(t,u)$,
in terms of which  (\ref{tequation}) becomes 
\beq \label{t-rho}
t\left( 1+\frac {9u^{10} \zeta^2}{t^2}\right) = \left\{t\left( 1-\frac {9u^{10} \zeta^2}{t^2}\right)^{4/3}
+u\left( 1-\frac {9u^{10} \zeta^2}{t^2}\right)^{2/3}+6 u^5\right\}\zeta,
\eeq
from which we can immediately conclude that
\beq \label{rho-omega}
\zeta=\frac{t}{t+u+6 u^5} +  \omega(t,u),
\eeq
where an expansion of $\omega$ in $u$ starts at order 10.
Any desired order can be obtained by further expanding $\omega$
\beq \label{omega_expansion}
\omega = u^{10}\sum_{n=0}^\infty a_n(t) u^n,
\eeq 
with the co-efficients $a_n(t)$ to be determined.
Putting  (\ref{omega_expansion}) and (\ref{rho-omega})
into (\ref{t-rho}), and equating co-efficients of powers 
of $u$, gives an iterative procedure for evaluating the co-efficients $a_n$
which can then be used to show that:
\begin{equation}
\omega= 7\frac{u^{10}}{t^{2}}-16\frac{u^{11}}{t^{3}}+27\frac{u^{12}}{t^{4}}-40\frac{u^{13}}{t^{5}}+55\frac{u^{14}}{t^{6}}-36\,{\frac {2+3\,{t}^{4}}{{t}^{7}}}
{u}^{15}
+O\left( {u}^{16} \right).
\end{equation}
Finally using this expansion in (\ref{p-z}) gives the required virial
expansion
\bea
p&=&\frac{2t u}{3}-{\frac {1}{3}}{u}^{2}+2\,{u}^{6}
-u^{10}\left(8\frac{u}{t}-9\frac{u^2}{t^2}+10\frac{u^3}{t^3}-
11\frac{u^4}{t^4}+12\frac{u^5}{t^5}\right.\\ 
& & \left.\hskip 6cm -13\,{\frac {1+6\,{t}^{4}}{{t}^{6}}}{u}^{6}\right)
+O
 \left( {u}^{17} \right),\nonumber
\eea
the first three terms of which are used in the text.

\pagebreak

\includegraphics[width=300pt,height=300pt]{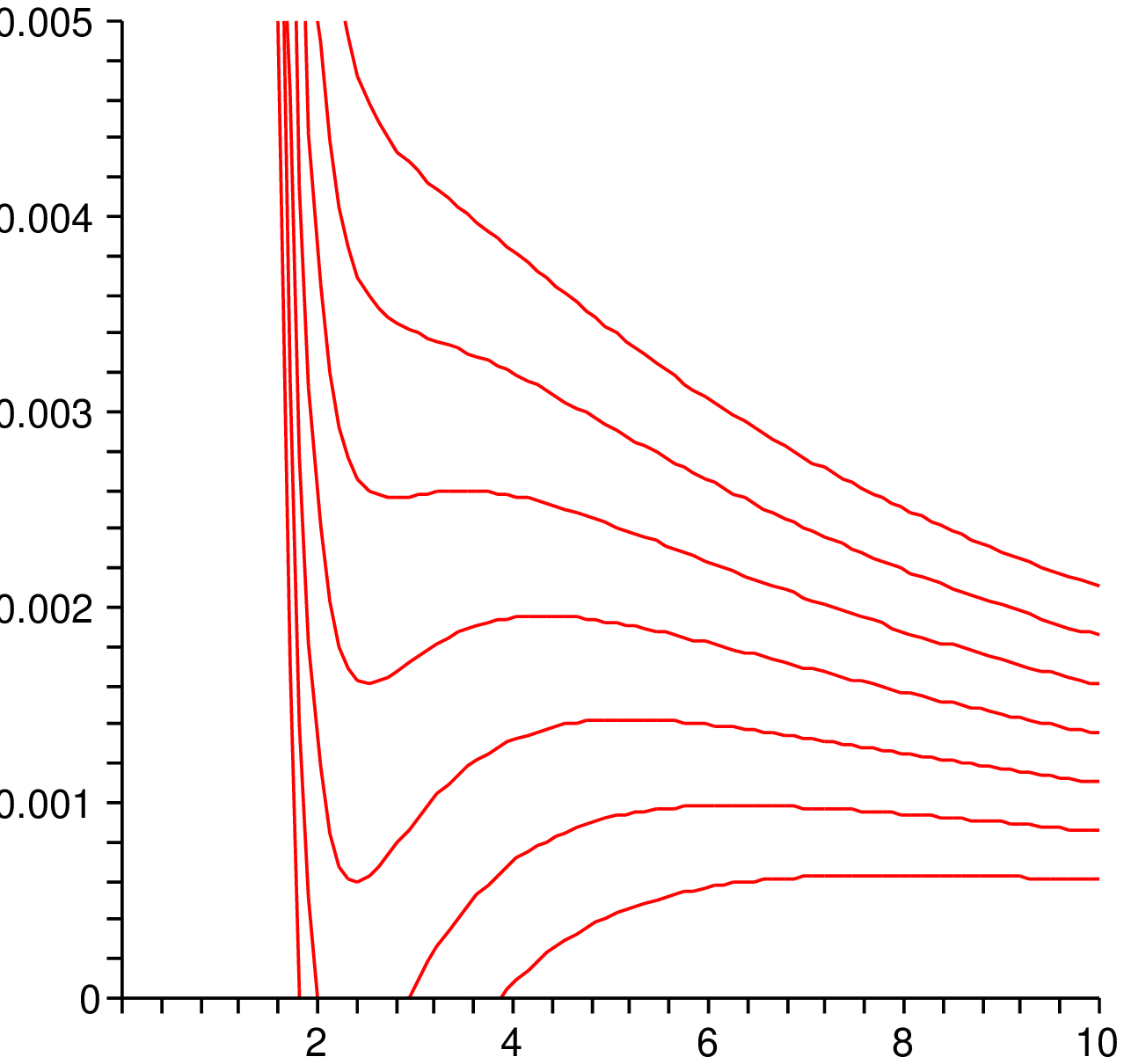}\par
\vskip 1cm
\centerline{Figure 1: $P$-$V$ diagram for $J=1$ and $Q=0$,
plotted using the approximation}
\leftline {in equation (\ref{PVapprox}).
$P$ is plotted as a function of $v^{\frac 1 3}$ 
for $T=0.02$, $0.025$, $0.3$,}
\centerline{ $0.35$, $0.04$, $0.045$ and $0.05$. The critical point for this value of $J$ is $T_c\approx 0.0413$,}
\centerline{$P_c\approx 0.00280$ and $v_c\approx 3.08$. $J>0$ causes the rapid rise in $P$ at low values of $v$.}

\pagebreak 
\includegraphics[width=300pt,height=300pt]{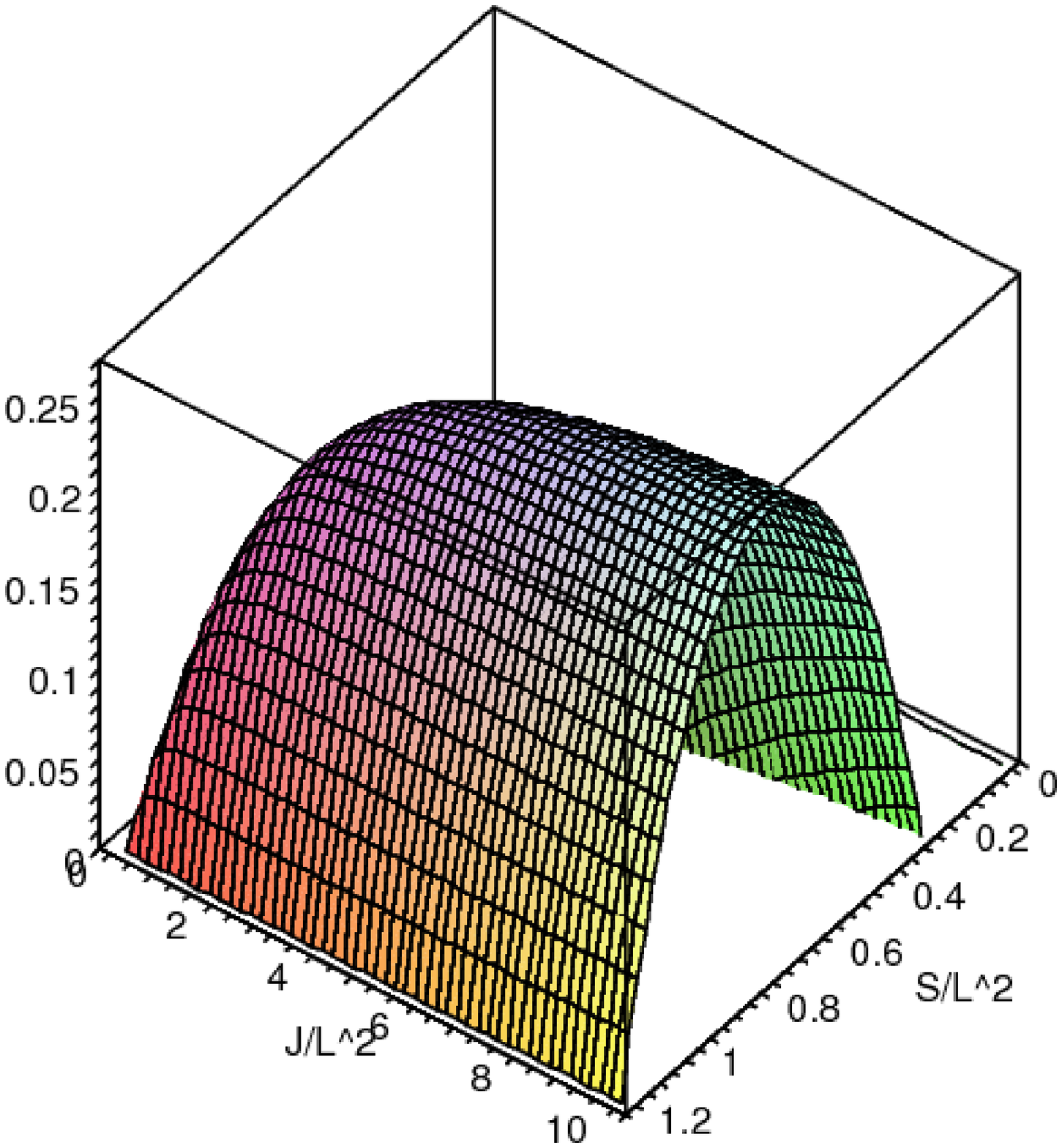}\par
\vskip 1cm
\centerline{Figure 2: $C_V$ as a function of $S/L^2$ and
$J/L^2$, with fixed volume set to $L^3$.}

\pagebreak
\includegraphics[width=300pt,height=300pt]{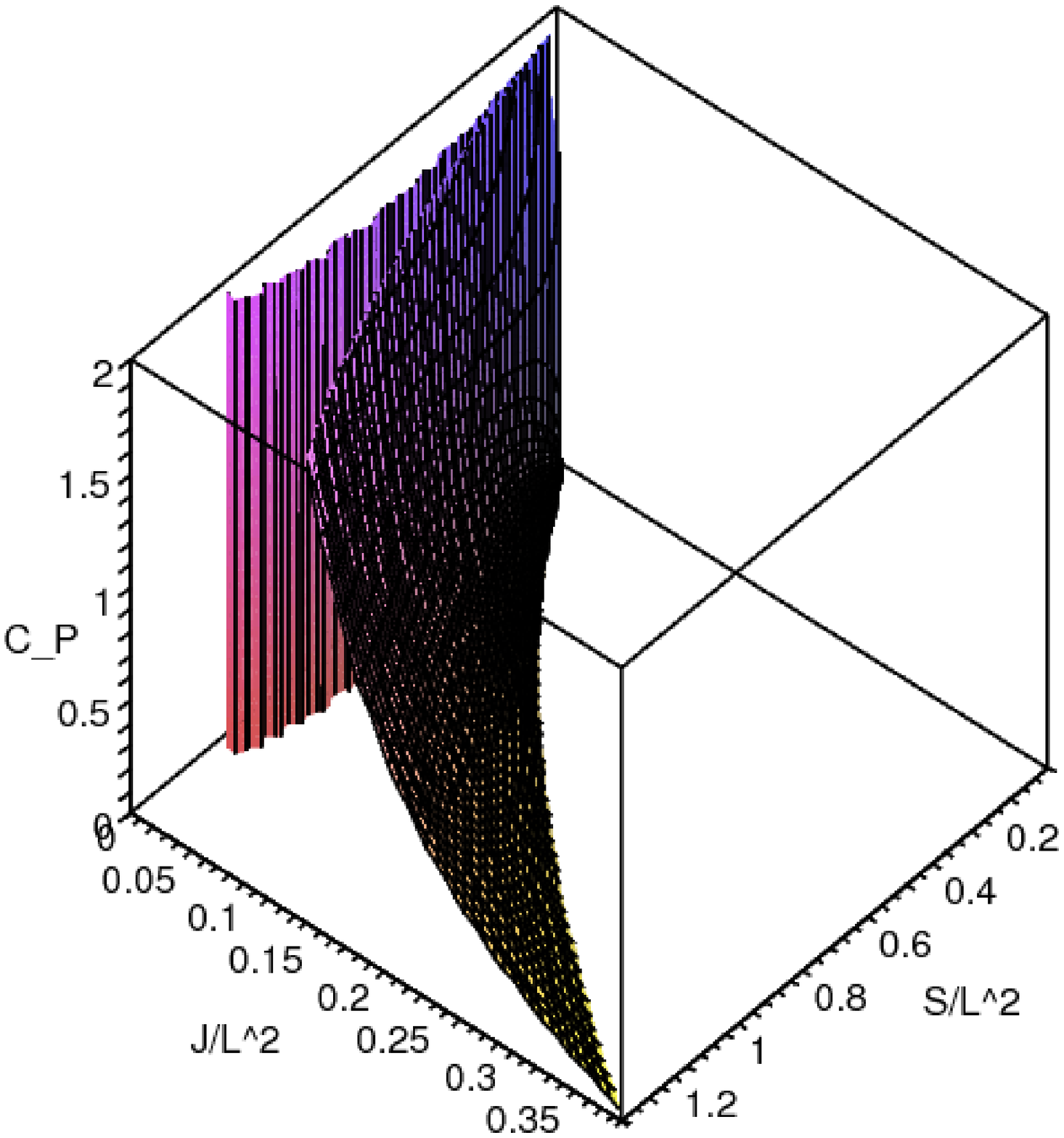}\par
\vskip 1cm
\centerline{Figure 3: $C_P$ as a function of $S/L^2$ and
$J/L^2$, with fixed pressure}
\centerline{ set to $\frac{3}{8\pi}$, {\it i.e} $L=1$.
$C_P$ diverges along curve IV in figure 4 and}
\centerline{vanishes along curve I.}
\pagebreak

\hskip -80pt 
\includegraphics[width=500pt,height=300pt]{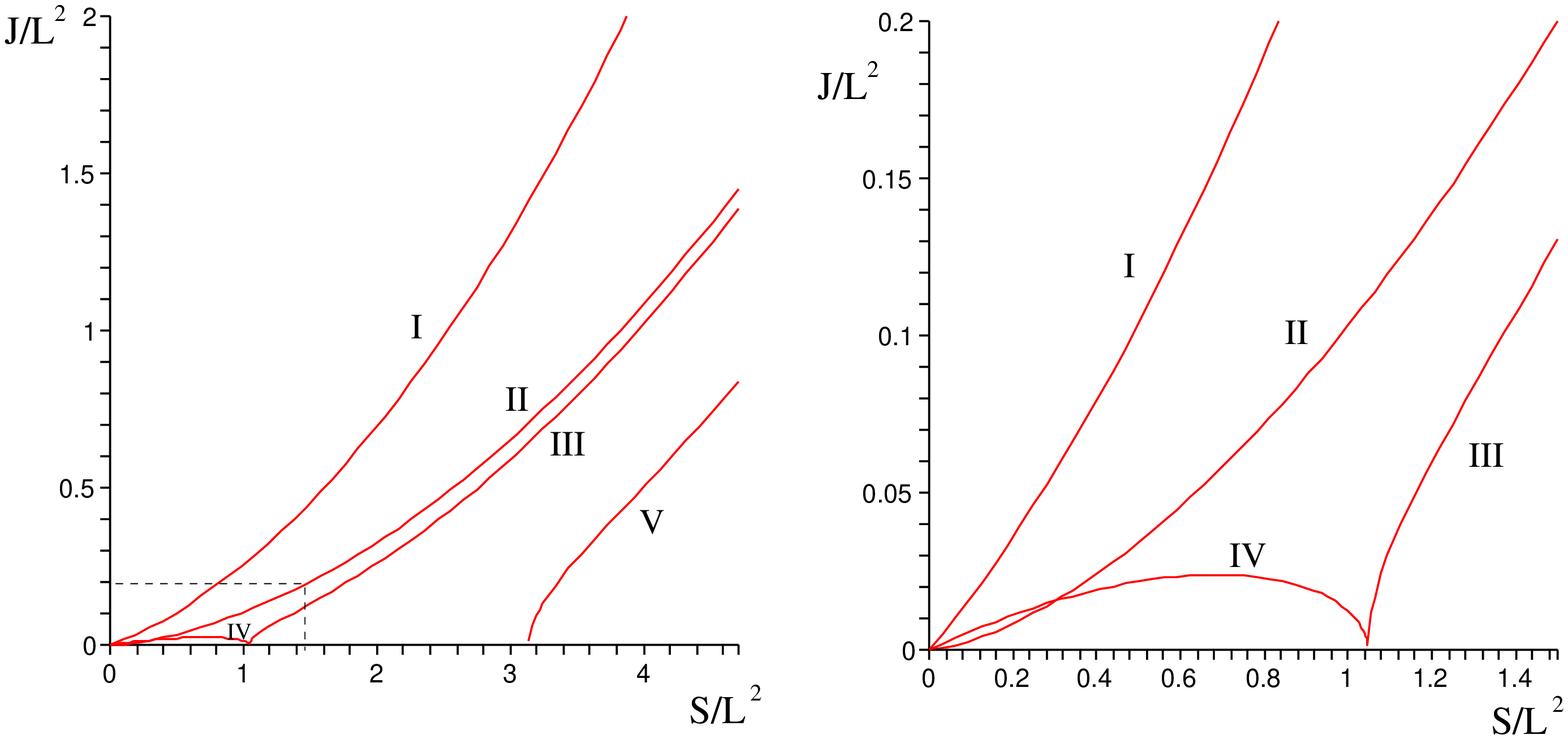}\par

\leftline{Figure 4: phase diagram for $q=0$, plotted in terms of $S/L^2$ and $J/L^2$.}
\leftline{The region above I is forbidden,
because $T<0$;}
\leftline{in the region above curve II the 3-d Einstein universe at infinity rotates}
\leftline{faster than the speed of light;} 
\leftline{curve III bounds the region of local stability (an analysis of the Gibbs free} 
\leftline{energy shows that the black hole is locally unstable above curve III);}
\leftline{$C_P$ diverges on curve IV;}
\leftline{in the region above curve V the black hole is unstable due
to the Hawking-}
\leftline{Page  phase transition;}
\leftline{in region below curve V the black hole is stable.}


\end{document}